\documentclass[11pt]{article}
\usepackage[margin=1in]{geometry}
\usepackage{authblk}
\usepackage{graphicx}
\usepackage{url}
\usepackage{natbib}
\usepackage{amsmath}

\title{GeoThinneR: An R Package for Efficient Spatial Thinning of Species Occurrences and Point Data}

\author[1,2,*]{Jorge Mestre-Tomás}

\affil[1]{Institute of Marine Sciences (ICM-CSIC), Renewable Marine Resources Department, Barcelona, Spain}
\affil[2]{Universitat Politècnica de València (UPV), Department of Applied Statistics and Operational Research, and Quality, Valencia, Spain}
\affil[*]{Corresponding author: \texttt{jorgemestre@icm.csic.es}}

\date{}

\begin{document}

\maketitle

\begin{center}
    \textbf{Note.} This is a preprint.
\end{center}

\begin{abstract}
\noindent
In this paper we present GeoThinneR, an R package for efficient and flexible spatial thinning of species occurrence data. Spatial thinning is a widely used preprocessing step in species distribution modeling (SDM) that can help reduce sampling bias, but existing R implementations rely on brute-force algorithms that scale poorly with large datasets. GeoThinneR implements multiple thinning approaches, including ensuring a minimum distance between points, subsampling points on a grid, and filtering based on decimal precision. To handle large datasets, it introduces two optimized algorithms based on local kd-trees and adaptive neighbor estimation, which greatly reduce memory usage and execution time. Additional functionalities such as group-wise thinning and point prioritization are included to facilitate its use in SDM workflows. We here provide performance benchmarks using both simulated and real-world data to demonstrate substantial performance improvements over existing tools.
\end{abstract}

\vspace{1em}
\noindent\textbf{Keywords:} spatial thinning, sampling bias, species distribution modeling, occurrence data, kd-trees, R package.

\section{Introduction}

In recent years the availability of spatial biodiversity data has increased, with repositories such as the Global Biodiversity Information Facility (GBIF, \url{https://www.gbif.org}) and the Ocean Biodiversity Information System (OBIS, \url{https://obis.org/}) providing datasets containing hundreds of thousands of species occurrence records. Larger datasets provide new opportunities in ecological research and species distribution modeling (SDM), which are used to model the species geographic distributions and ecological niches based on occurrence records and environmental variables \citep{pulliam2000relationship, elith2009species, miller2010species, guisan2013predicting}. However, as dataset sizes grow, the computational requirements for processing and filtering them increase too. Species occurrence data are often affected by spatial sampling biases due to factors such as opportunistically collected data, uneven sampling efforts, or combining information from diverse sources resulting in heterogeneous datasets with substantial spatial sampling bias \citep{boakes2010distorted, lobo2011exploring, beck2014spatial, meyer2016range, hughes2021sampling}. These biases result in clusters of records that can artificially increase spatial autocorrelation among occurrences and cause the model to overfit the environmental conditions in these oversampled regions, leading to models that reflect patterns in sampling effort rather than true species distributions \citep{wisz2008effects, phillips2009sample, higa2015mapping, cosentino2021geographic, steen2021spatial}.

Several approaches exist to address sampling bias \citep{inman2021comparing, baker2024effective, moudry2024optimising}, including adjusting background points \citep{barber2022target, vollering2019bunching}, applying environmental filtering \citep{castellanos2019environmental}, incorporating bias-related covariates \citep{varela2014environmental, chauvier2021novel}, or modeling preferential sampling \citep{diggle2010geostatistical, amaral2024model}. One of the most widely used methods to mitigate these biases across the geographic space is spatial thinning, which selectively filters points based on specified criteria to reduce the overrepresentation of the densest locations \citep{veloz2009spatially, boria2014spatial}. The two most common thinning methods in SDM are distance-based thinning, where points are removed if they fall within a specified minimum distance of another point \citep{mi2023global}, and grid-based thinning, which retains only a specified number of points per grid cell \citep{melton2022climatic}.

While existing distance-based thinning tools in R, such as \texttt{spThin} \citep{aiello2015spthin} and \texttt{enmSdmX} \citep{smith2023including}, are widely used, they have several limitations that are particularly aggravated as dataset sizes continue to increase. These tools rely on brute force algorithms for nearest-neighbor searches given a distance threshold, which are unfeasible to run on personal computers as they scale poorly for datasets containing hundreds of thousands of points. On the other hand, grid-based thinning, such as the one implemented in the \texttt{dismo} package \citep{dismoRpackage}, offers better computational efficiency but lacks precision when requiring a specified distance threshold. Additionally, these tools have poor functionality and usability as they often focus on a single thinning approach and lack advanced customization features that could optimize thinning for specific SDM applications, such as thinning occurrences across multiple species \citep{noori2024window}, retaining an exact number of occurrences \citep{steen2021spatial}, or prioritizing points based on data features \citep{yu2023future}.

To overcome these challenges, we introduce \texttt{GeoThinneR} \citep{geothinner2025}, an R package designed to provide efficient and flexible spatial thinning for large-scale datasets. The package is available from the Comprehensive R Archive Network (CRAN) at \url{https://cran.r-project.org/package=GeoThinneR}. \texttt{GeoThinneR} integrates multiple thinning approaches into a single package simplifying user experience (distance, grid, and decimal precision-based thinning), offers an optimized distance-based thinning using enhanced algorithms based on \emph{kd}-tree structures, which improve nearest-neighbor search efficiency \citep{friedman1975algorithm}, together with additional specific functionalities for SDM workflows including thinning by species or groups, the ability to retain an exact number of occurrences and prioritization of points based on user-defined variables.

In this paper, we present \texttt{GeoThinneR} as a valuable tool for researchers working with large-scale species occurrence data, combining computational efficiency with enhanced usability and SDM-specific functionalities. In Section~\ref{packagedescription}{2}, we present the methods and features implemented in \texttt{GeoThinneR}. Section~\ref{modifiedthinning}{3} contains information about the optimized algorithms for large datasets together with a performance comparison between the distance-based methods available. Section~\ref{performancethinning}{4} benchmarks the performance of the package with existing software, both with simulated and real-world datasets. Finally, Section~\ref{conclusions}{5} presents a short conclusion of the work and discusses future directions.

\section{GeoThinneR package description}\label{packagedescription}

Given a set of points \(\mathbf{S} = \{ \mathbf{x}_1, \mathbf{x}_2, \dots, \mathbf{x}_n \}\), spatial thinning aims to filter \(\mathbf{S}\) to obtain a subset \(\mathbf{S}' \subseteq \mathbf{S}\) that meets a specified selection criteria. Different thinning methods can be employed to apply this criteria, such as enforcing a minimum separation distance between retained points, limiting the number of points per grid cell, or rounding coordinates to a specified precision. The thinning process iteratively evaluates each point, referred to as the query point \(q\), identifies neighbor points within \(\mathbf{S}\), and applies the selection criteria to return a thinned subset \(\mathbf{S}'\).

The \texttt{GeoThinneR} package is implemented in R and can be downloaded from the Comprehensive R Archive Network (CRAN) at \url{https://cran.r-project.org/package=GeoThinneR}. Comprehensive documentation and usage examples are available on GitHub and CRAN. One of the advantages of \texttt{GeoThinneR} is its flexibility and ease of use, allowing users to efficiently apply different spatial thinning configurations to their datasets. In this section, we demonstrate how to use the package and explore its various thinning methods and additional functionalities. To illustrate the basic usage, we simulate a toy dataset of 100 uniformly distributed points for two species across an area of 1 x 1 degree and apply spatial thinning using the \texttt{thin\_points()} function:

\begin{verbatim}
library(GeoThinneR)
# Simulate data
set.seed(2547)
sim_data <- data.frame(
  lon = runif(100, 0, 1),
  lat = runif(100, 0, 1),
  group = sample(c("species_1", "species_2"), 100, replace = TRUE)
)

thin_results <- thin_points(sim_data,
  method = "distance", thin_dist = 10,
  trials = 10, all_trials = TRUE, seed = 567
)
\end{verbatim}

\texttt{GeoThinneR} returns the results structured as a \texttt{GeoThinned} object, which contains: a list of logical vectors indicating which points were retained in each trial (\texttt{x\$retained}), the method used for thinning (\texttt{x\$method}), the parameters used for thinning (\texttt{x\$params}), and the original dataset (\texttt{x\$original\_data}).

The \texttt{summary()} function provides information about the thinning process, including the number of points retained in each trial, the nearest neighbor distance (NND), and the coverage of the retained points.

\begin{verbatim}
summary(thin_results)
\end{verbatim}

\begin{verbatim}
#> Summary of GeoThinneR Results
#> -----------------------------
#> Method used       : distance 
#> Distance metric   : haversine 
#> 
#> Number of points:
#>   Original               100
#>   Thinned                 46
#>   Retention ratio      0.460
#> 
#> Nearest Neighbor Distances [km]
#>             Min 1st Qu. Median   Mean 3rd Qu.    Max
#> Original  1.054   3.699  5.349  5.917   7.474 16.400
#> Thinned  10.152  10.942 11.856 12.624  14.261 17.031
#> 
#> Spatial Coverage (Convex Hull Area) [km2]
#>   Original         10453.599
#>   Thinned          10142.351
\end{verbatim}

The output indicates that we retained 46\% of the points and that the minimum distance between points has increased without reducing too much the spatial coverage or extent of the points compared to the original dataset. Using the \texttt{plot()} function, we can visualize the original and thinned dataset.

\begin{verbatim}
# Visualize largest thinned trial
plot(thin_results)
# Show retained points only
plot(thin_results, show_original = FALSE)
# Change the colors
plot(thin_results, col_original = "red", col_thinned = "blue")
\end{verbatim}

While determining the maximum number of points to retain that meet the thinning constraint is simple and straightforward for grid-based and precision-based thinning, for distance-based thinning is a non-deterministic problem in areas such as combinatorial mathematics know as the set packing problem. For so, similarly to \texttt{spThin}, \texttt{GeoThinneR} allows users to perform multiple independent thinning trials, that will yield varying-size sets of unthinned points, returning either all attempts or the largest retained subset that meets the thinning criteria. Using the \texttt{largest()} and \texttt{get\_trial()} functions, users can extract the largest thinned dataset or any of the trials, respectively.

\begin{verbatim}
# Extract the largest thinned dataset
largest(thin_results)
# Extract the index of the largest dataset
largest_index(thin_results)
# Extract the second trial
get_trial(thin_results, 2)
\end{verbatim}

\subsection{Overview of thinning methods}\label{overview-of-thinning-methods}

\begin{figure}
\includegraphics[width=1\linewidth]{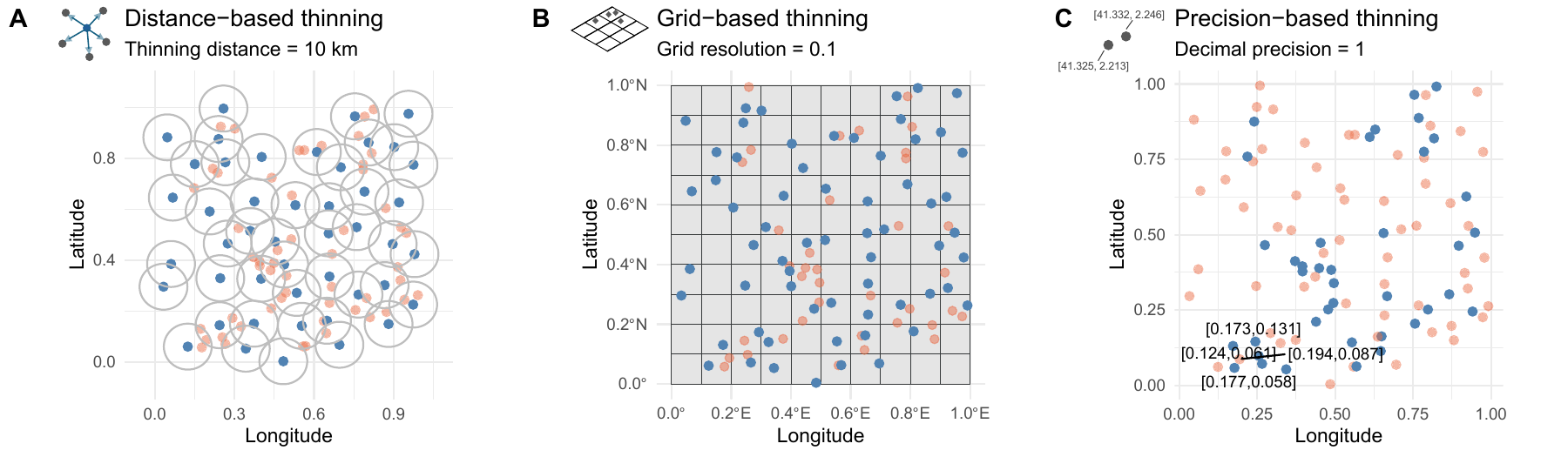} \caption{Spatial thinning examples using the three methods implemented in GeoThinneR: (A) distance-based thinning with a thinning distance of 10 km, (B) grid-based thinning with a grid resolution of 0.1 degrees and retaining one point per grid cell, and (C) precision-based thinning with coordinates rounded to one decimal place. Blue points indicate retained points and red points those removed during the thinning process.}\label{fig:thinmethods}
\end{figure}

\texttt{GeoThinneR} integrates three distinct methods to provide flexibility: distance-based, grid-based, and precision-based thinning (Figure \ref{fig:thinmethods}). For better usability, all methods are implemented within a modular wrapper function, \texttt{thin\_points()}, which includes the thinning methods with optimized algorithms for large datasets.

Distance-based thinning (\texttt{method="distance"}) ensures that all retained points are separated by at least a user-defined minimum distance \(d\) \citep{aiello2015spthin}. However, efficiently identifying neighbor points becomes computationally challenging as dataset sizes increase. For a given set of points \(\mathbf{S}\), the algorithm iteratively removes points with neighbors within \(d\) based on either Haversine (geographic coordinates) or Euclidean (Cartesian coordinates) distance. By default, \texttt{GeoThinneR} uses the Haversine formula \citep{chopde2013landmark} to compute great-circle distances:
\begin{equation}
D(\mathbf{x}_i, \mathbf{x}_j) = 2R \cdot \arcsin \left( \sqrt{\sin^2 \left( \frac{\text{lat}_j - \text{lat}_i}{2} \right) + \cos(\text{lat}_j) \cdot \cos(\text{lat}_i) \cdot \sin^2 \left( \frac{\text{lon}_j - \text{lon}_i}{2} \right)} \right),
  \label{eq:haversine}
\end{equation}
where \(R\) is the radius of the Earth (by default, 6371 km), and \((\text{lon}, \text{lat})\) are the longitude and latitude coordinates of \(\mathbf{x}_i\) and \(\mathbf{x}_j\). The Haversine formula assumes a smooth spherical Earth, providing sufficient accuracy for most spatial analyses \citep{maria2020measure}. For example, to obtain points separated at least by 10 km we could use a similar code as before:

\begin{verbatim}
# Distance-based thinning
dist_thin <- thin_points(sim_data, method="distance", thin_dist=10, seed=8237)
\end{verbatim}

Grid-based thinning (\texttt{method="grid"}) is an alternative that stratifies points \(\mathbf{S}\) into equal-sized spatial grid cells \(\mathcal{G} = \{ G_1, G_2, \dots, G_m \}\), selecting a subset of points \(\mathbf{S}'_{G_j} \subseteq \{ \mathbf{x}_i \in \mathbf{S} : \mathbf{x}_i \in G_j \}\) from each cell \(G_j \in \mathcal{G}\) \citep{dismoRpackage}. This method is computationally more efficient than distance-based thinning as it does not need to compute neighbor distances, making it suitable for large datasets where strict separation distances are unnecessary.

We can provide the grid size in kilometers (\texttt{thin\_dist}), the resolution of the grid (\texttt{resolution}), or a raster object to use as grid template (\texttt{raster\_obj}).

\begin{verbatim}
# Grid-based thinning using resolution
grid_thin <- thin_points(sim_data, method="grid", resolution=0.1, seed=9137)

# Using raster layer
rast_obj <- terra::rast(xmin = 0, xmax = 1, ymin = 0, ymax = 1, res = 0.1)
grid_thin <- thin_points(sim_data, method="grid", raster_obj=rast_obj,seed=54898)
\end{verbatim}

Precision-based thinning (\texttt{method="precision"}) removes duplicate points by rounding coordinates to a specified decimal precision, which is a very common procedure when dealing with georeferenced data in SDM \citep{kindt2023climate}. This method is particularly useful when dealing with heterogeneous datasets with varying levels of spatial precision or when reducing density without enforcing a strict minimum distance is sufficient. Since it avoids distance calculations entirely, precision-based thinning is particularly fast and scales efficiently with dataset size.

\begin{verbatim}
# Distance-based thinning
prec_thin <- thin_points(sim_data, method="precision", precision=1, seed=5674)
\end{verbatim}

\subsection{Optimization for large datasets}\label{optimization-for-large-datasets}

As the size of spatial datasets increases, finding exact nearest neighbors by scanning the complete set \(\mathbf{S}\) and computing all pairwise distances for each query point \(q\) becomes computationally unfeasible. This makes nearest neighbor searches one of the primary bottlenecks in spatial thinning methods with exact distance thresholds. To address this, \texttt{GeoThinneR} implements four different search strategies:

\begin{itemize}
\item
  \texttt{search\_type="brute"} is a greedy algorithm similar to the one implemented in \texttt{spThin} and \texttt{enmSdmX}, which calculates all pairwise distances between points using the \texttt{fields} package \citep{fieldsRpackage}. While being the most straightforward approach, it scales poorly with large datasets due to its \(O(n^2)\) complexity.
\item
  \texttt{search\_type="kd\_tree"} uses \emph{kd}-tree structures from the \texttt{nabor} package \citep{elsebergcomparison}, which were proposed to theoretically reduce nearest-neighbor searches time complexity to \(O(\log n)\) \citep{friedman1975algorithm}. Although \emph{kd}-trees can significantly improve performance for large datasets, they may suffer from the ``curse of dimensionality'' or in cases where exhaustive searches are required to identify exact nearest neighbors, their performance can be equivalent to or worse than brute-force algorithms \citep{ram2019revisiting}.
\item
  \texttt{search\_type="local\_kd\_tree"} is one of the minor modifications we propose to the \emph{kd}-trees search algorithm to reduce run time and memory usage for large-scale datasets. To enhance scalability, instead of creating a single huge \emph{kd}-tree where lots of queries have to be evaluated, \texttt{GeoThinneR} subdivides the spatial domain into multiple subregions, each with an independent and smaller \emph{kd}-tree. This partitioning strategy largely reduces memory usage by building smaller and more manageable search trees.
\item
  \texttt{search\_type="k\_estimation"} is an alternative approach to improve \emph{kd}-tree performance. When searching for nearest-neighbors within a given distance, the exact number of neighbors \(k\) per query point \(q\) is unknown, so the algorithm evaluates whether all points in the tree lie within the search radius bound from \(q\). To reduce the number of searches performed in the tree for each query point \(q\), we propose a method to estimate the maximum possible number of neighbors \(k_\text{max}\). This estimation reduces the number of searches per point from \(n\) to \(k_\text{max}\), reducing unnecessary computations of distant points while still finding the exact number of neighbors.
\end{itemize}

These optimizations significantly improve computational efficiency, making \texttt{GeoThinneR} suitable for large-scale SDM applications. The underlying algorithms and their performance trade-offs are further detailed in Section~\ref{modifiedthinning}{3}.

\subsection{Additional functionalities}\label{additional-functionalities}

\begin{figure}
\includegraphics[width=1\linewidth]{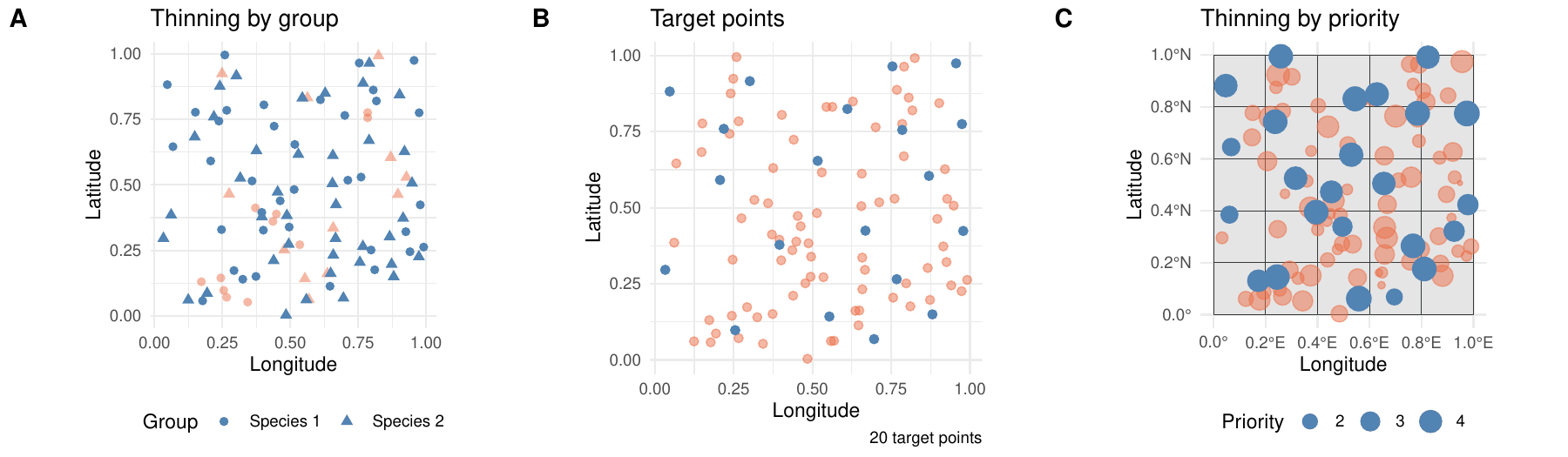} \caption{Additional spatial thinning functionalities available in GeoThinneR: (A) group-wise thinning applied independently for each species (B) retaining a fixed number of points (20 points in this example) while ensuring spatial separation, and (C) prioritizing points based on larger variable values within each grid cell. Retained points are shown in blue, and removed points are in red.}\label{fig:addfeat}
\end{figure}

In addition to its core thinning methods, \texttt{GeoThinneR} includes several functionalities designed for SDM workflows that users may require when running their analysis. Firstly, \texttt{GeoThinneR} allows to perform group-specific thinning within a single dataset. This is useful when working with datasets containing multiple groups, such as different species or presence/absence data, where thinning needs to be applied independently within each group (Figure \ref{fig:addfeat}a).

\begin{verbatim}
# Thinning by group
precision_thin_group <- thin_points(sim_data,
  method = "precision", precision = 1,
  group_col = "group", seed = 2948
)
\end{verbatim}

Secondly, in some cases, users may need to retain a fixed number of points while maintaining spatial representativeness, such as for class balancing in presence/absence datasets. \texttt{GeoThinneR} allows users to specify a target number of retained points and attempts to retain the specified number of points while guaranteeing a minimum distance threshold between them (Figure \ref{fig:addfeat}b). The process for retaining points selects the most spatially distant points to maximize geographical representativeness. This approach is supported only via the brute-force method as it requires all pairwise distance computations to select most distant points.

\begin{verbatim}
# Thinning target points
target_thin <- thin_points(sim_data, target_points=20, thin_dist=10, seed=675)
\end{verbatim}

Finally, when applying the thinning algorithm and having to choose which points to remove and retain, users can specify variables that influence which points are preferentially retained, making it possible to prioritize records based on ecological relevance, data precision, or uncertainty (Figure \ref{fig:addfeat}c). For example, heterogeneous datasets from sources such as GBIF often include records of varying spatial precision. Prioritization allows users to retain records with higher accuracy or filter occurrences based on both geographical and environmental importance. This feature is compatible with both grid-based and precision-based thinning, retaining those points with the highest priority from each set of neighboring points.

\begin{verbatim}
# Thinning priority
sim_data$priority <- runif(100, 1, 5)
rast_obj <- terra::rast(xmin = 0, xmax = 1, ymin = 0, ymax = 1, res = 0.2)
priority_thin <- thin_points(sim_data,
  method = "grid", raster_obj = rast_obj,
  priority = sim_data$priority, seed = 3455
)
\end{verbatim}

\section{Modified distance-based thinning algorithm based on kd-tree}\label{modifiedthinning}

\begin{figure}
\includegraphics[width=1\linewidth]{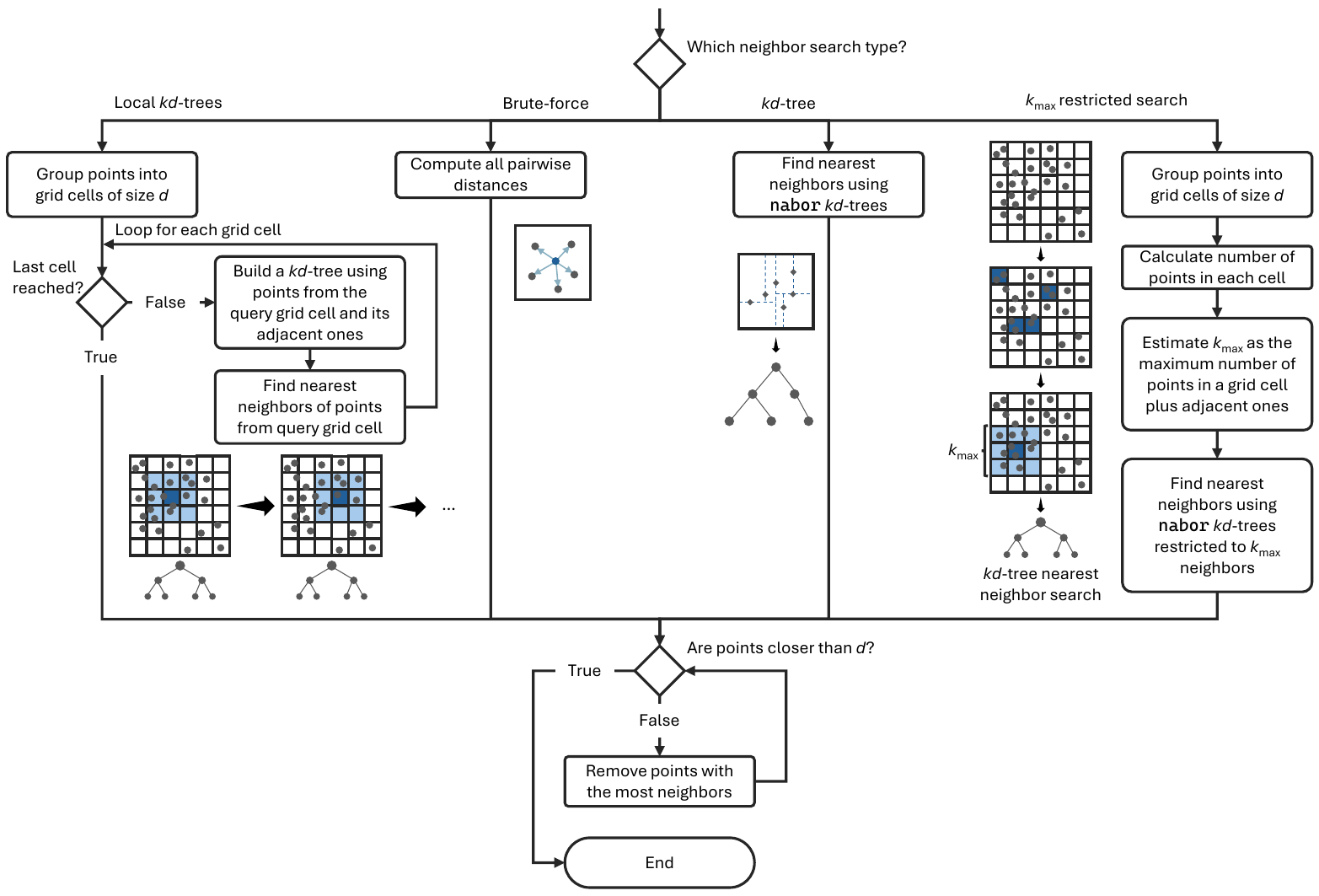} \caption{Diagram illustrating the four nearest neighbor search strategies implemented in GeoThinneR for distance-based thinning. The methods include brute-force, standard kd-trees, local kd-trees, and restricted kd-trees using estimated k-max. Each approach computes neighboring points within a thinning distance before proceeding with point removal.}\label{fig:distthinalg}
\end{figure}

The distance-based thinning process can be conceptualized as a two-step procedure: (1) identifying neighboring points within a given distance threshold, and (2) selectively removing points to meet the selection criteria. Current methods in SDM workflows use greedy approaches in the first step, computing all pairwise distances between points and resulting in a time complexity of \(O(n^2)\) and high memory usage. \texttt{GeoThinneR}, like \texttt{spThin} \citep{aiello2015spthin} and \texttt{enmSdmX} \citep{smith2023including}, implements a brute-force method using the \texttt{rdist.earth()} function from the \texttt{fields} package \citep{fieldsRpackage}. To improve performance, we propose two minor modifications to \emph{kd}-tree nearest-neighbor search algorithms improving scalability for large datasets (Figure \ref{fig:distthinalg}). These methods significantly reduce computational complexity while ensuring exact thinning distances.

\subsection{Traditional kd-tree nearest neighbor search}\label{traditional-kd-tree-nearest-neighbor-search}

A \emph{kd}-tree \citep{friedman1975algorithm} is a data structure that organizes points in a \(k\)-dimensional space by recursively partitioning the data along hyperplanes creating a binary search tree. Many \emph{kd}-tree variants exist, but the essence is that they recursively partition the space containing the set of points \(\mathbf{S}\) into small hyper-rectangles that contain a small subset \(\mathbf{S}'_i \subset \mathbf{S}\) of the input points \citep{panigrahy2008improved}. Each node represents a \(k\)-dimensional point, and each non-leaf node is a splitting hyperplane that divides the space into two partitions \citep{ram2019revisiting}.

\texttt{GeoThinneR} implements \emph{kd}-tree construction and nearest-neighbor searches using the \texttt{nabor} package, which wraps the \texttt{libnabo} library \citep{elsebergcomparison}. Since these \emph{kd}-trees are built on Euclidean distances, geographic coordinates (latitude and longitude) are transformed into Cartesian coordinates \((x, y, z)\) before constructing the binary tree:
\begin{equation}
\begin{split}
x &= R \cdot \cos(\text{lat}) \cdot \cos(\text{lon}) \\
y &= R \cdot \cos(\text{lat}) \cdot \sin(\text{lon}) \\
z &= R \cdot \sin(\text{lat})
\end{split}
  \label{eq:cartesian}
\end{equation}
where \(\text{lat}\) and \(\text{lon}\) are the latitude and longitude in radians, respectively (\(\frac{\pi}{180}(\text{lon}, \text{lat})\)), and \(R\) is the Earth's radius. This results in a three-dimensional \emph{kd}-tree.

However, performance may be worsened in higher dimensions or when many points must be evaluated as potential neighbors of the query point \(q\). The time required to find the neighbors of is affected by the search for candidate neighbors and the time it takes to evaluate among all candidates which ones are actually neighbors \citep{ram2019revisiting}. While approximate searches can reduce computational time by tuning the precision-recall trade-off, exact searches in large datasets require exhaustive searches, leading to higher computational complexity To improve scalability, we introduce two additional optimizations: local \emph{kd}-trees partitioning and restricted neighbor searches.

\subsection{Local kd-trees for scalable searches}\label{local-kd-trees-for-scalable-searches}

To reduce the complexity of searching within a single \emph{kd}-tree constructed for the entire set of points \(\mathbf{S}\), which requires evaluating all points to identify neighbors within the distance threshold, we propose constructing local \emph{kd}-trees by subdividing the spatial domain into multiple smaller regions, each with its own tree. Although creating multiple \emph{kd}-trees is time-consuming, memory usage and overall runtime are improved for large datasets, as neighbor searches are less exhaustive when evaluating smaller subsets \(\mathbf{S}'_i \subset \mathbf{S}\).

To find exact neighbors using this approach, we first divide the spatial domain into a grid with cell sizes equal to the thinning distance. We use geohash coordinates to efficiently find points within each grid cell. Then, for each grid cell, we identify points within the cell and its adjacent cells, as these are potential neighbors. Recursively, an independent \emph{kd}-tree is constructed for each grid cell using the \texttt{nabor} package, and neighbors are identified for each query point within the cell.

\noindent\textbf{Algorithm: Identifying neighbors using local \emph{kd}-trees}

\begin{verbatim}
Input: coordinates, a matrix of longitude-latitude points;
       thin_dist, thinning distance in kilometers

Steps:
  1. Convert geographic coordinates to Cartesian coordinates
  2. Group points into grid cells of size thin_dist
  3. for each unique grid cell do
     a. Identify all points within the current cell (cell_ids)
     b. Identify neighboring cells and their points (neighbor_ids)
     c. Combine cell_ids and neighbor_ids
     d. Build a local kd-tree using the combined points
     e. For each point in cell_ids find neighbors
        within thin_dist using nabor::knn()
     f. Remove self-references and store neighbor indices
  4. end for
  5. return: A list of neighbor indices for all points
\end{verbatim}

This strategy maintains computational efficiency while avoiding memory overloads of constructing and querying large trees. Furthermore, local \emph{kd}-trees can be easily executed in parallel, providing a straightforward way to scale thinning operations across large datasets.

\subsection{Restricted neighbor searches}\label{restricted-neighbor-searches}

When performing the nearest-neighbor searches using \emph{kd}-trees from \texttt{nabor}, we have to specify the number of neighbors to find \(k\) which increases or decreases the search complexity. As we don't know the number of neighbors each point \(q\) will have we set this value \(k\) to the set \(\mathbf{S}\) size \(n\), meaning all points are checked to determine whether they are neighbors of the query point. However, for large datasets, this approach introduces unnecessary computational costs by evaluating distant points. Instead, we propose an adaptive neighbor search that estimates \(k_{\text{max}}\) based on local point densities to reduce the number of nearest neighbors to find for each query points.

The process involves dividing the spatial domain into grid cells of size equal to the thinning distance \(d\), identifying the densest cells, and calculating the maximum number of neighbors a point \(k_{\text{max}}\) can have based on local point densities. Once \(k_{\text{max}}\) is estimated, it is used to limit the number of neighbors evaluated during the \emph{kd}-tree search.

\noindent\textbf{Algorithm: Restricted neighbor searches using estimated \(k_{\text{max}}\)}

\begin{verbatim}
Input: coordinates, a matrix of longitude-latitude points;
       thin_dist, thinning distance in kilometers;
       k_max, maximum number of nearest neighbors to find

Steps:
  1. Estimate k_max:
     a. Define cell_size = thin_dist / 111.32
     b. Create grid cells based on cell_size
     c. Count number of points in each grid cell
     d. Identify the densest grid cells and adjacent cells
     e. Calculate k_max as the maximum number of points within 
        the densest cells and their neighbors
  2. Compute neighbors:
     a. Build a kd-tree using the nabor package
     b. For each point, find up to k_max neighbors within thin_dist
     c. Remove self-references from neighbor lists
  3. return: A list of neighbor indices for all points
\end{verbatim}

This method significantly reduces computational time, particularly in datasets where the maximum number of neighbors per point is much smaller than the total number of points. However, one limitation is that densely populated areas or large thinning distances may inflate \(k_{\text{max}}\), making the method less effective. In such cases, the local \emph{kd}-tree method with parallelization provides a suitable alternative.

\subsection{Performance comparison of neighbor search methods}\label{performancesearchtype}

In this section, we evaluate the computational efficiency of the different neighbor search methods implemented in \texttt{GeoThinneR} using simulated datasets of varying sizes and spatial structures. All tests were conducted using R 4.3.3 on a Windows 11 computer with 128 GB of RAM and an Intel(R) Xeon(R) Gold 6240R processor featuring 24 cores, 48 threads, and a base clock speed of 2.40 GHz. The benchmark code is available in the supplementary R files.

We simulated datasets within a unit square domain \([0,1] \times [0,1]\), ranging from 1,000 to 50,000 points, under three spatial processes to mimic different real-world scenarios: (i) a homogeneous Poisson process simulating complete spatial randomly distributed points, (ii) a Matern clustered point process generated using the \texttt{rMatClust} function from the \texttt{spatstat} package \citep{spatstatRpackage} with 10 cluster centers (\texttt{kappa}), a cluster radius of 0.15 (\texttt{scale}), and a varying number of points per cluster (\texttt{mu}), and (iii) a mixed spatial pattern dataset combining the randomly distributed points and the clustered processes. Finally, we benchmarked the methods using two thinning distances (1 km and 10 km) to highlight the impact of increasing the number of points that need to be removed and how each method scales with different search complexities.

We compared the neighbor search methods in \texttt{GeoThinneR} using execution time (seconds) and peak RAM usage (MB) to evaluate the computational speed and memory requirements, which are critical for large-scale spatial datasets. Each method was tested three times, and the mean values were calculated.

\begin{figure}
\includegraphics[width=1\linewidth]{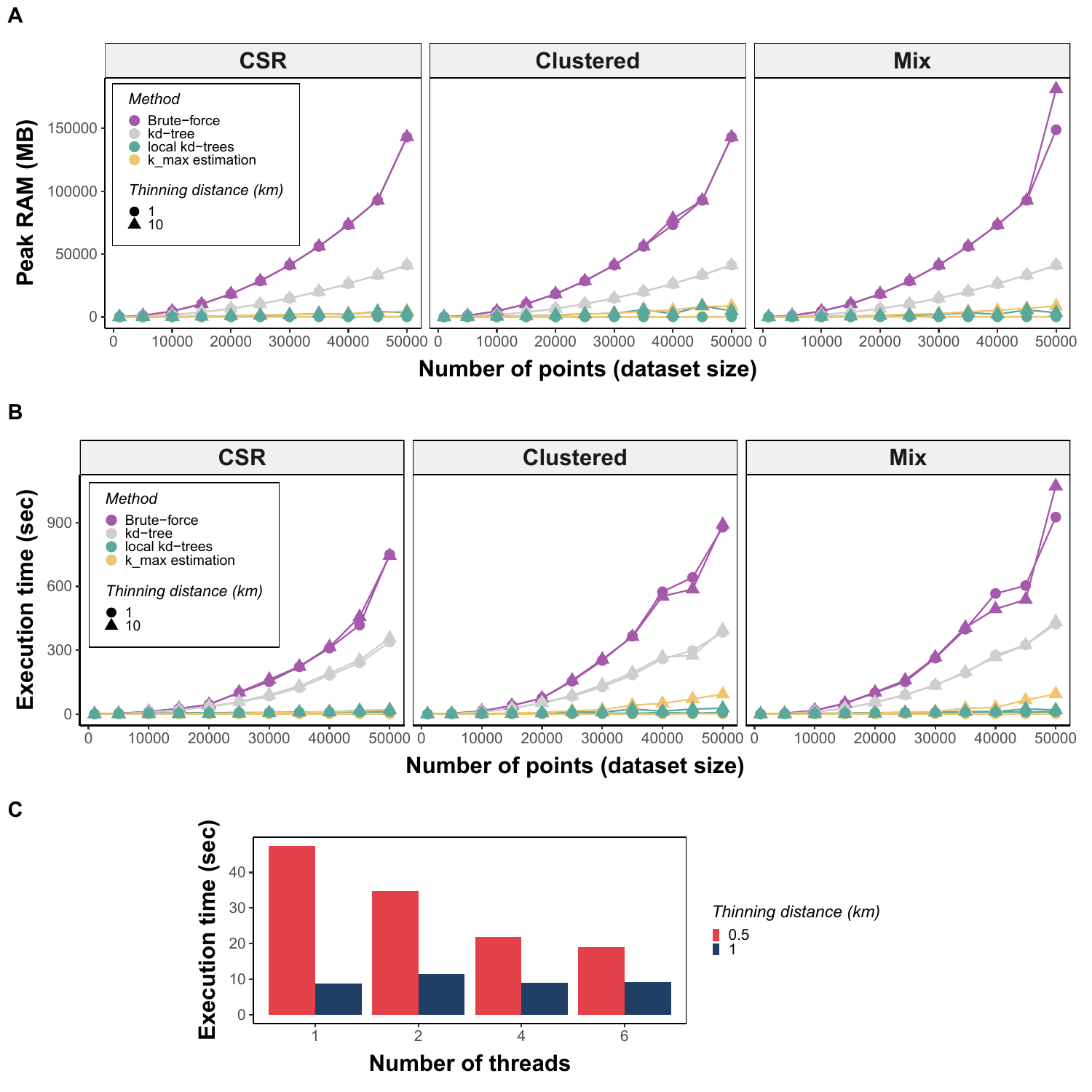} \caption{Performance benchmark for neighbor searches implemented in GeoThinneR distance-based thinning. (A) Peak RAM usage (MB) and (B) execution time (seconds) across three spatial data types (CSR, clustered, and mixed spatial pattern) for dataset sizes ranging from 1,000 to 50,000 points. (C) Performance comparison of local kd-trees with parallelization for 50,000 randomly distributed points using two thinning distances (0.5 km and 1 km). The optimized algorithms (local kd-trees and k-estimation) scale better and provide superior computational efficiency than brute-force and standard kd-tree, especially for large datasets.}\label{fig:benchmarksearchtype}
\end{figure}

The two modified algorithms (\(k_\text{max}\) estimation and local \emph{kd}-trees) consistently outperform the brute-force and traditional \emph{kd}-tree methods in both peak memory usage (Figure \ref{fig:benchmarksearchtype}a) and execution time (Figure \ref{fig:benchmarksearchtype}b) across all spatial data types. The traditional \emph{kd}-tree method provides better performance compared to the brute-force method for small data sets, however, for larger data sets, memory usage and execution time increase due to exhaustive searches in deeper trees to find all neighbors of each point. However, the two optimized methods are orders of magnitude faster and less memory-intensive. For small thinning distances, the \(k_\text{max}\) estimation algorithm shows the best performance in terms of both time and memory usage, because the estimated number of neighbors \(k_\text{max}\) remains low, reducing unnecessary searches. However, at larger thinning distances, and particularly for the clustered and mix datasets, as the dataset size increases, this algorithm goes slower and uses more memory than the local \emph{kd}-trees approach. This is because the algorithm estimates a high value for \(k_\text{max}\) and, therefore, increases search complexity.

Alternatively, the local \emph{kd}-tree approach, which generates an independent \emph{kd}-tree for each subregion, offers improved memory efficiency and outperforms the other methods as thinning distances increase, as fewer trees need to be built. This method can also be executed in parallel, which can further improve run time for large datasets, especially when many partitions are created due to a small thinning distance or a large spatial extent (Figure \ref{fig:benchmarksearchtype}c). However, if the number of partitions is not very large, as in the case of the 1 km thinning distance in this example, parallel execution may not bring any improvement.

\section{Performance benchmark}\label{performancethinning}

In this section, we compare the performance of \texttt{GeoThinneR} with two widely used R packages for spatial thinning: \texttt{spThin} and \texttt{enmSdmX}. We compare the execution time and memory usage of the distance-based thinning algorithms available in \texttt{GeoThinneR} version 2.0.0 against the \texttt{thin()} function from \texttt{spThin} version 0.2.0 and the \texttt{geoThin()} function from \texttt{enmSdmX} version 1.2.10. In Section~\ref{benchsim}{4.1}, we compare the performance using simulated data, and in Section~\ref{benchreal}{4.2}, we benchmark the methods using a real-world large dataset.

\begin{verbatim}
brute <- thin_points(data,
  method = "distance", search_type = "brute",
  thin_dist = thin_dist, trials = trials
)
kd_tree <- thin_points(data,
  method = "distance", search_type = "kd_tree",
  thin_dist = thin_dist, trials = trials
)
local_kd_tree <- thin_points(data,
  method = "distance", search_type = "local_kd_tree",
  thin_dist = thin_dist, trials = trials
)
k_estimation <- thin_points(data,
  method = "distance", search_type = "k_estimation",
  thin_dist = thin_dist, trials = trials
)
spThin <- thin(data,
  lat.col = "lat", long.col = "lon", spec.col = "group",
  thin.par = thin_dist, reps = trials, locs.thinned.list.return = TRUE,
  write.files = FALSE, write.log.file = FALSE, verbose = FALSE
)
enmSdmX <- geoThin(data, 
  minDist = thin_dist * 1000, longLat = c("lon", "lat"), method = "complete"
)
\end{verbatim}

\subsection{Benchmark with simulated data}\label{benchsim}

The benchmark tests were performed on the complete thinning workflow using the simulated datasets described in Section~\ref{performancesearchtype}{3.4}. Each thinning method was executed with a single trial to obtain the maximum number of retained points per run, and each test was repeated three times to ensure reliable timing estimates. Here, we compared all methods across three spatial data types (random, clustered, and mixed) using again two thinning distances of 1 and 10 kilometers.

\begin{figure}
\includegraphics[width=1\linewidth]{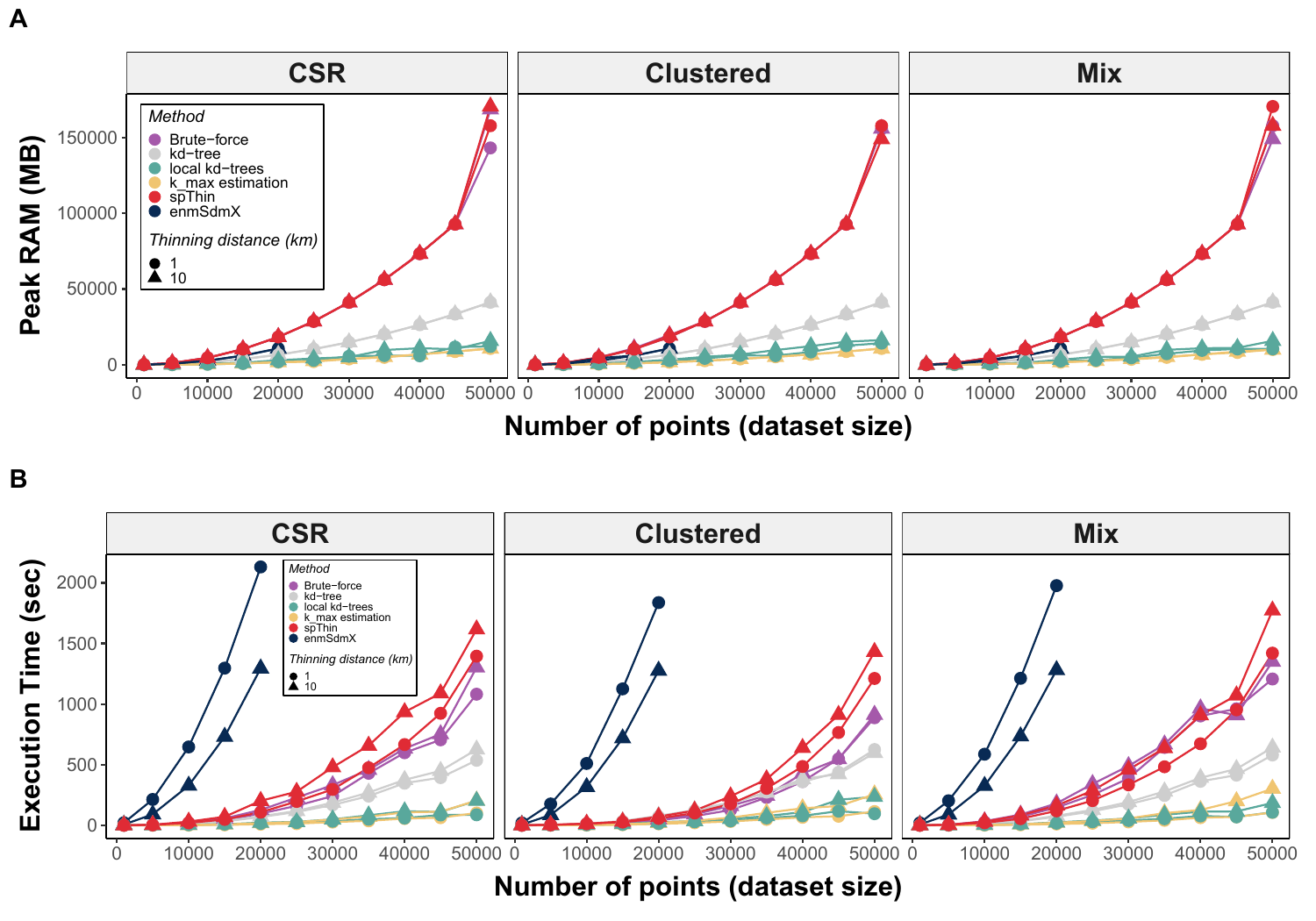} \caption{Performance benchmark for distance-based thinning across GeoThinneR, spThin, and enmSdmX. (A) Peak RAM usage (MB) and (B) execution time (seconds) across three spatial data types for varying dataset sizes. The enmSdmX method was only evaluated up to 20,000 points due to its high computational cost on larger datasets. The optimized methods of GeoThinneR (local kd-trees and k-estimation) offer better memory and runtime performance across all settings.}\label{fig:benchmarkthinning}
\end{figure}

The optimized algorithms implemented in \texttt{GeoThinneR} demonstrate significant improvements in both memory efficiency (Figure \ref{fig:benchmarkthinning}a) and execution time (Figure \ref{fig:benchmarkthinning}b) compared to existing tools. The \texttt{enmSdmX} package exhibited the longest execution times, making it unsuitable for large datasets, so we limited its tests to 20,000 points. \texttt{spThin}, although it's more efficient, shows similar performance to the brute-force method from \texttt{GeoThinneR}. In contrast, our new modified methods demonstrate better scalability in both runtime and peak memory usage, making them more practical for large-scale spatial thinning.

The local \emph{kd}-tree and \(k_\text{max}\) estimation algorithms in \texttt{GeoThinneR} present the best scalability for large datasets. Patterns are similar to the ones when compared the neighbor searches methods. The \(k_\text{max}\) estimation method is best suited for small thinning distances and without relatively dense regions, while the local \emph{kd}-tree method is offers a good alternative for highly clustered datasets or large thinning distances. This flexibility ensures that \texttt{GeoThinneR} can adapt to various real-world SDM workflows more efficiently than existing tools

\subsection{Application to real-world data}\label{benchreal}

To illustrate the performance of \texttt{GeoThinneR} on real-world datasets, we applied its optimized thinning methods and compared them with \texttt{spThin}. We used a small dataset of 8,340 occurrence records for the loggerhead sea turtle (\emph{Caretta caretta}) in the Mediterranean Sea, and a large and high clustered dataset of 80,163 points for the yellowfin tuna (\emph{Thunnus albacares}) collected globally from 1950 to 2025. Both datasets were downloaded from GBIF \citep{GBIFCaretta} and are very heterogeneous including data from different sources, locations, and sampling methods. The data are available in \texttt{GeoThinneR} package and can be loaded using the \texttt{data()} function.

\begin{verbatim}
# Loggerhead sea turtle
data(caretta)
# Yellowfin tuna
data(thunnus)
\end{verbatim}

Here we show the performance of the local \emph{kd}-tree and \(k_\text{max}\) estimation algorithms implemented in \texttt{GeoThinneR}, comparing their computational performance against \texttt{spThin} using varying thinning distances. The objective is to show the trade-offs, best- and worst-case scenarios for each method, that's why we use varying thinning distances and dataset sizes. So we are not trying to identify the optimal thinning distance for SDM applications, this comparison highlights the scalability and efficiency of the algorithms. It is important to note that choosing an appropriate thinning distance for SDM is challenging and often requires to be tuned empirically \citep{veloz2009spatially, soley2019sufficient}.

\begin{figure}
\includegraphics[width=1\linewidth]{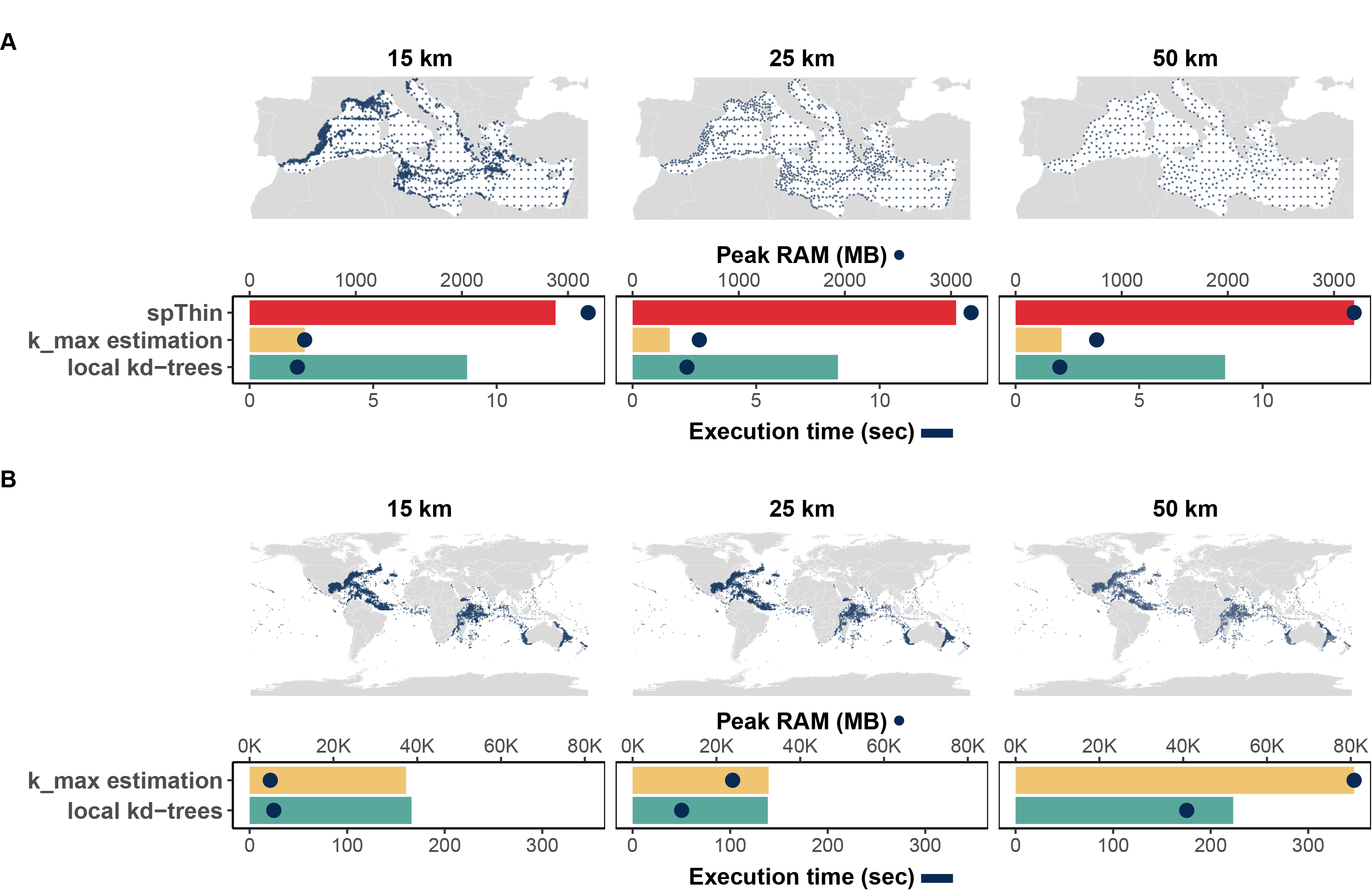} \caption{Execution time (bars) and memory usage (points) of GeoThinneR and spThin across varying thinning distances for occurrence data of (A) loggerhead turtles and (B) yellowfin tuna from GBIF. For the yellowfin tuna dataset, the results for spThin are not shown because the run could not be completed due to excessive memory usage.}\label{fig:benchmarkrealdata}
\end{figure}

The results, shown in Figure \ref{fig:benchmarkrealdata}, show significant performance differences even for the smaller dataset. \texttt{GeoThinneR} not only executed faster but also demonstrated substantially lower memory usage, addressing one of the key limitations of current spatial thinning methods in SDM (Figure \ref{fig:benchmarkrealdata}a). The \(k_\text{max}\) estimation algorithm was particularly efficient, with run times between one and two seconds due to low neighbor density that minimizes search complexity.

In contrast, the yellowfin tuna large dataset presented a different scenario (Figure \ref{fig:benchmarkrealdata}b). \texttt{spThin} failed to execute due to excessive memory demands, surpassing our 128 GB RAM limit after several minutes, which highlight the scalability challenges of brute-force methods. Both algorithms in \texttt{GeoThinneR} maintained low memory usage, even at this dataset size. However, the \(k_\text{max}\) estimation method experienced slower performance as thinning distances increased because the dataset presents highly clustered ares which inflate \(k_\text{max}\) estimates and increase search complexity. To address this, the local \emph{kd}-tree approach, executed here with multithreading (6 threads), proved both faster and less memory-intensive, being the method with best performance in this context. These results illustrate that while \(k_\text{max}\) estimation is highly efficient for moderately sized or uniformly distributed datasets, the local \emph{kd}-tree method with parallelization offers a memory efficient solution for large and dense datasets.

Here we present the code used to generate the thinned datasets represented in Figure \ref{fig:benchmarkrealdata}:

\begin{verbatim}
caretta_thin <- thin_points(
  caretta,
  lon_col = "decimalLongitude",
  lat_col = "decimalLatitude",
  method = "distance",
  thin_dist = thin_dist, # One of 10, 25, 50
  search_type = "k_estimation",
  trials = 10
)

thunnus_thin <- thin_points(
  thunnus,
  lon_col = "decimalLongitude",
  lat_col = "decimalLatitude",
  method = "distance",
  thin_dist = thin_dist, # One of 10, 25, 50
  search_type = "local_kd_tree",
  trials = 1,
  n_cores = 6
)
\end{verbatim}

\section{Conclusions}\label{conclusions}

The \texttt{GeoThinneR} R package offers a flexible and scalable implementation of spatial thinning methods, addressing key limitations of existing tools. It integrates multiple thinning approaches (distance-, grid-, and precision-based) into a single wrapper function, enhancing usability. Also, by incorporating optimized \emph{kd}-tree algorithms for distance-based thinning, such as local \emph{kd}-tree partitioning and adaptive \(k_\text{max}\) estimation, the package significantly reduces execution time and memory usage for large datasets. Additionally, \texttt{GeoThinneR} includes customization features designed for species distribution modeling workflows, such as group-wise thinning and point prioritization. While future developments should be made in improving the computational performance of distance-based methods, incorporating additional thinning approaches, and estimating an optimal thinning distance, \texttt{GeoThinneR} represents a step forward in improving the efficiency of spatial thinning methods in SDM.

\section*{Acknowledgments}\label{acknowledgments}

This package has been developed as part of the ProOceans (PID2020-118097RB-I00) and SOSPen (PID2021-124831OA-I00) projects funded by the Spanish Ministry of Science and Innovation. The authors also acknowledge the institutional support of the ``Severo Ochoa Center of Excellence'' accreditation (CEX2019-000928-S) to the Institute of Marine Science (ICM-CSIC). Additionally, this work is part of the Integrated Marine Ecosystem Assessments (iMARES) research group, funded by the Agència de Gestió d'Ajuts Universitaris i de Recerca of the Generalitat de Catalunya (2021 SGR 00435).

\section*{Supplementary materials}\label{supplementary-materials}

Supplementary materials are available in addition to this article and can be accessed via Zenodo (\url{https://doi.org/10.5281/zenodo.15356817}):

\begin{itemize}
\item
  \emph{GeoThinneR\_replication\_small.R}: reproduces examples and figures from Sections 2 and 3. Includes a short benchmark and demonstrations of package features using small datasets.
\item
  \emph{GeoThinneR\_replication\_large.R}: reproduces figures and results from Section 4. Includes a larger benchmark to evaluate performance of GeoThinneR and other packages using simulated and real-world datasets.
\item
  \emph{caretta\_download.R}: downloads occurrence data for loggerhead turtles (\emph{Caretta caretta}) from GBIF.
\item
  \emph{thunnus\_download.R}: downloads occurrence data for yellowfin tuna (\emph{Thunnus albacares}) from GBIF.
\end{itemize}


\bibliographystyle{apalike}
\bibliography{GeoThinneR_preprint}

\begin{thebibliography}{}

\bibitem[Aiello-Lammens et~al., 2015]{aiello2015spthin}
Aiello-Lammens, M.~E., Boria, R.~A., Radosavljevic, A., Vilela, B., and Anderson, R.~P. (2015).
\newblock {spThin}: An r package for spatial thinning of species occurrence records for use in ecological niche models.
\newblock {\em Ecography}, 38(5):541--545.

\bibitem[Amaral et~al., 2024]{amaral2024model}
Amaral, A. V.~R., Krainski, E.~T., Zhong, R., and Moraga, P. (2024).
\newblock Model-based geostatistics under spatially varying preferential sampling.
\newblock {\em Journal of Agricultural, Biological and Environmental Statistics}, 29(4):766--792.

\bibitem[Baddeley et~al., 2015]{spatstatRpackage}
Baddeley, A., Rubak, E., and Turner, R. (2015).
\newblock {\em Spatial Point Patterns: Methodology and Applications with R}.
\newblock Chapman and Hall/CRC Press, London.

\bibitem[Baker et~al., 2024]{baker2024effective}
Baker, D.~J., Maclean, I.~M., and Gaston, K.~J. (2024).
\newblock Effective strategies for correcting spatial sampling bias in species distribution models without independent test data.
\newblock {\em Diversity and Distributions}, 30(3):e13802.

\bibitem[Barber et~al., 2022]{barber2022target}
Barber, R.~A., Ball, S.~G., Morris, R.~K., and Gilbert, F. (2022).
\newblock Target-group backgrounds prove effective at correcting sampling bias in maxent models.
\newblock {\em Diversity and Distributions}, 28(1):128--141.

\bibitem[Beck et~al., 2014]{beck2014spatial}
Beck, J., B{\"o}ller, M., Erhardt, A., and Schwanghart, W. (2014).
\newblock Spatial bias in the {GBIF} database and its effect on modeling species' geographic distributions.
\newblock {\em Ecological Informatics}, 19:10--15.

\bibitem[Boakes et~al., 2010]{boakes2010distorted}
Boakes, E.~H., McGowan, P.~J., Fuller, R.~A., Chang-qing, D., Clark, N.~E., O'Connor, K., and Mace, G.~M. (2010).
\newblock Distorted views of biodiversity: Spatial and temporal bias in species occurrence data.
\newblock {\em PLoS biology}, 8(6):e1000385.

\bibitem[Boria et~al., 2014]{boria2014spatial}
Boria, R.~A., Olson, L.~E., Goodman, S.~M., and Anderson, R.~P. (2014).
\newblock Spatial filtering to reduce sampling bias can improve the performance of ecological niche models.
\newblock {\em Ecological Modelling}, 275:73--77.

\bibitem[Castellanos et~al., 2019]{castellanos2019environmental}
Castellanos, A.~A., Huntley, J.~W., Voelker, G., and Lawing, A.~M. (2019).
\newblock Environmental filtering improves ecological niche models across multiple scales.
\newblock {\em Methods in Ecology and Evolution}, 10(4):481--492.

\bibitem[Chauvier et~al., 2021]{chauvier2021novel}
Chauvier, Y., Zimmermann, N.~E., Poggiato, G., Bystrova, D., Brun, P., and Thuiller, W. (2021).
\newblock Novel methods to correct for observer and sampling bias in presence-only species distribution models.
\newblock {\em Global Ecology and Biogeography}, 30(11):2312--2325.

\bibitem[Chopde and Nichat, 2013]{chopde2013landmark}
Chopde, N.~R. and Nichat, M. (2013).
\newblock Landmark based shortest path detection by using a* and haversine formula.
\newblock {\em International Journal of Innovative Research in Computer and Communication Engineering}, 1(2):298--302.

\bibitem[Cosentino and Maiorano, 2021]{cosentino2021geographic}
Cosentino, F. and Maiorano, L. (2021).
\newblock Is geographic sampling bias representative of environmental space?
\newblock {\em Ecological Informatics}, 64:101369.

\bibitem[Diggle et~al., 2010]{diggle2010geostatistical}
Diggle, P.~J., Menezes, R., and Su, T.-l. (2010).
\newblock Geostatistical inference under preferential sampling.
\newblock {\em Journal of the Royal Statistical Society Series C: Applied Statistics}, 59(2):191--232.

\bibitem[{Douglas Nychka} et~al., 2021]{fieldsRpackage}
{Douglas Nychka}, {Reinhard Furrer}, {John Paige}, and {Stephan Sain} (2021).
\newblock {fields}: Tools for spatial data.
\newblock R package version 16.3.

\bibitem[Elith and Leathwick, 2009]{elith2009species}
Elith, J. and Leathwick, J.~R. (2009).
\newblock Species distribution models: Ecological explanation and prediction across space and time.
\newblock {\em Annual Review of Ecology, Evolution, and Systematics}, 40(1):677--697.

\bibitem[Elseberg et~al., 2012]{elsebergcomparison}
Elseberg, J., Magnenat, S., Siegwart, R., and N{\"u}chter, A. (2012).
\newblock Comparison of nearest-neighbor-search strategies and implementations for efficient shape registration.
\newblock {\em Journal of Software Engineering for Robotics (JOSER)}, 3(1):2--12.

\bibitem[Friedman et~al., 1975]{friedman1975algorithm}
Friedman, J.~H., Bentley, J.~L., and Finkel, R.~A. (1975).
\newblock {\em An Algorithm for Finding Best Matches in Logarithmic Time}.
\newblock Department of Computer Science, Stanford University.

\bibitem[{GBIF.org}, 2024]{GBIFCaretta}
{GBIF.org} (2024).
\newblock Occurrence download.

\bibitem[Guisan et~al., 2013]{guisan2013predicting}
Guisan, A., Tingley, R., Baumgartner, J.~B., Naujokaitis-Lewis, I., Sutcliffe, P.~R., Tulloch, A.~I., Regan, T.~J., Brotons, L., McDonald-Madden, E., Mantyka-Pringle, C., et~al. (2013).
\newblock Predicting species distributions for conservation decisions.
\newblock {\em Ecology Letters}, 16(12):1424--1435.

\bibitem[Higa et~al., 2015]{higa2015mapping}
Higa, M., Yamaura, Y., Koizumi, I., Yabuhara, Y., Senzaki, M., and Ono, S. (2015).
\newblock Mapping large-scale bird distributions using occupancy models and citizen data with spatially biased sampling effort.
\newblock {\em Diversity and Distributions}, 21(1):46--54.

\bibitem[Hijmans et~al., 2023]{dismoRpackage}
Hijmans, R.~J., Phillips, S., Leathwick, J., and Elith, J. (2023).
\newblock {\em {dismo}: Species Distribution Modeling}.
\newblock R package version 1.3-14.

\bibitem[Hughes et~al., 2021]{hughes2021sampling}
Hughes, A.~C., Orr, M.~C., Ma, K., Costello, M.~J., Waller, J., Provoost, P., Yang, Q., Zhu, C., and Qiao, H. (2021).
\newblock Sampling biases shape our view of the natural world.
\newblock {\em Ecography}, 44(9):1259--1269.

\bibitem[Inman et~al., 2021]{inman2021comparing}
Inman, R., Franklin, J., Esque, T., and Nussear, K. (2021).
\newblock Comparing sample bias correction methods for species distribution modeling using virtual species.
\newblock {\em Ecosphere}, 12(3):e03422.

\bibitem[Kindt et~al., 2023]{kindt2023climate}
Kindt, R., Abiyu, A., Borchardt, P., Dawson, I., Demissew, S., Graudal, L., Jamnadass, R., Lilles{\o}, J.-P., Moestrup, S., Pedercini, F., et~al. (2023).
\newblock {\em The Climate Change Atlas for Africa of Tree Species Prioritized for Forest Landscape Restoration in Ethiopia: A Description of Methods Used to Develop the Atlas}.
\newblock CIFOR.

\bibitem[Lobo and Tognelli, 2011]{lobo2011exploring}
Lobo, J.~M. and Tognelli, M.~F. (2011).
\newblock Exploring the effects of quantity and location of pseudo-absences and sampling biases on the performance of distribution models with limited point occurrence data.
\newblock {\em Journal for Nature Conservation}, 19(1):1--7.

\bibitem[Maria et~al., 2020]{maria2020measure}
Maria, E., Budiman, E., Taruk, M., et~al. (2020).
\newblock Measure distance locating nearest public facilities using haversine and euclidean methods.
\newblock In {\em Journal of Physics: Conference Series}, volume 1450, page 012080. IOP Publishing.

\bibitem[Melton et~al., 2022]{melton2022climatic}
Melton, A.~E., Clinton, M.~H., Wasoff, D.~N., Lu, L., Hu, H., Chen, Z., Ma, K., Soltis, D.~E., and Soltis, P.~S. (2022).
\newblock Climatic niche comparisons of eastern north american and eastern asian disjunct plant genera.
\newblock {\em Global Ecology and Biogeography}, 31(7):1290--1302.

\bibitem[Mestre-Tomás, 2025]{geothinner2025}
Mestre-Tomás, J. (2025).
\newblock {\em {GeoThinneR}: Efficient Spatial Thinning of Species Occurrences}.
\newblock R package version 2.0.0.

\bibitem[Meyer et~al., 2016]{meyer2016range}
Meyer, C., Jetz, W., Guralnick, R.~P., Fritz, S.~A., and Kreft, H. (2016).
\newblock Range geometry and socio-economics dominate species-level biases in occurrence information.
\newblock {\em Global Ecology and Biogeography}, 25(10):1181--1193.

\bibitem[Mi et~al., 2023]{mi2023global}
Mi, C., Ma, L., Yang, M., Li, X., Meiri, S., Roll, U., Oskyrko, O., Pincheira-Donoso, D., Harvey, L.~P., Jablonski, D., et~al. (2023).
\newblock Global protected areas as refuges for amphibians and reptiles under climate change.
\newblock {\em Nature Communications}, 14(1):1389.

\bibitem[Miller, 2010]{miller2010species}
Miller, J. (2010).
\newblock Species distribution modeling.
\newblock {\em Geography Compass}, 4(6):490--509.

\bibitem[Moudr{\`y} et~al., 2024]{moudry2024optimising}
Moudr{\`y}, V., Bazzichetto, M., Remelgado, R., Devillers, R., Lenoir, J., Mateo, R.~G., Lembrechts, J.~J., Sillero, N., Lecours, V., Cord, A.~F., et~al. (2024).
\newblock Optimising occurrence data in species distribution models: Sample size, positional uncertainty, and sampling bias matter.
\newblock {\em Ecography}, 2024(12):e07294.

\bibitem[Noori et~al., 2024]{noori2024window}
Noori, S., Hofmann, A., R{\"o}dder, D., Husemann, M., and Rajaei, H. (2024).
\newblock A window to the future: Effects of climate change on the distribution patterns of iranian zygaenidae and their host plants.
\newblock {\em Biodiversity and Conservation}, 33(2):579--602.

\bibitem[Panigrahy, 2008]{panigrahy2008improved}
Panigrahy, R. (2008).
\newblock An improved algorithm finding nearest neighbor using kd-trees.
\newblock In {\em Latin American Symposium on Theoretical Informatics}, pages 387--398. Springer.

\bibitem[Phillips et~al., 2009]{phillips2009sample}
Phillips, S.~J., Dud{\'\i}k, M., Elith, J., Graham, C.~H., Lehmann, A., Leathwick, J., and Ferrier, S. (2009).
\newblock Sample selection bias and presence-only distribution models: Implications for background and pseudo-absence data.
\newblock {\em Ecological Applications}, 19(1):181--197.

\bibitem[Pulliam, 2000]{pulliam2000relationship}
Pulliam, H.~R. (2000).
\newblock On the relationship between niche and distribution.
\newblock {\em Ecology Letters}, 3(4):349--361.

\bibitem[Ram and Sinha, 2019]{ram2019revisiting}
Ram, P. and Sinha, K. (2019).
\newblock Revisiting kd-tree for nearest neighbor search.
\newblock In {\em Proceedings of the 25th ACM SIGKDD International Conference on Knowledge Discovery \& Data Mining}, pages 1378--1388.

\bibitem[Smith et~al., 2023]{smith2023including}
Smith, A.~B., Murphy, S.~J., Henderson, D., and Erickson, K.~D. (2023).
\newblock Including imprecisely georeferenced specimens improves accuracy of species distribution models and estimates of niche breadth.
\newblock {\em Global Ecology and Biogeography}, 32(3):342--355.

\bibitem[Soley-Guardia et~al., 2019]{soley2019sufficient}
Soley-Guardia, M., Carnaval, A.~C., and Anderson, R.~P. (2019).
\newblock Sufficient versus optimal climatic stability during the late quaternary: Using environmental quality to guide phylogeographic inferences in a neotropical montane system.
\newblock {\em Journal of Mammalogy}, 100(6):1783--1807.

\bibitem[Steen et~al., 2021]{steen2021spatial}
Steen, V.~A., Tingley, M.~W., Paton, P.~W., and Elphick, C.~S. (2021).
\newblock Spatial thinning and class balancing: Key choices lead to variation in the performance of species distribution models with citizen science data.
\newblock {\em Methods in Ecology and Evolution}, 12(2):216--226.

\bibitem[Varela et~al., 2014]{varela2014environmental}
Varela, S., Anderson, R.~P., Garc{\'\i}a-Vald{\'e}s, R., and Fern{\'a}ndez-Gonz{\'a}lez, F. (2014).
\newblock Environmental filters reduce the effects of sampling bias and improve predictions of ecological niche models.
\newblock {\em Ecography}, 37(11):1084--1091.

\bibitem[Veloz, 2009]{veloz2009spatially}
Veloz, S.~D. (2009).
\newblock Spatially autocorrelated sampling falsely inflates measures of accuracy for presence-only niche models.
\newblock {\em Journal of Biogeography}, 36(12):2290--2299.

\bibitem[Vollering et~al., 2019]{vollering2019bunching}
Vollering, J., Halvorsen, R., Auestad, I., and Rydgren, K. (2019).
\newblock Bunching up the background betters bias in species distribution models.
\newblock {\em Ecography}, 42(10):1717--1727.

\bibitem[Wisz et~al., 2008]{wisz2008effects}
Wisz, M.~S., Hijmans, R., Li, J., Peterson, A.~T., Graham, C., Guisan, A., and Group, N. P. S. D.~W. (2008).
\newblock Effects of sample size on the performance of species distribution models.
\newblock {\em Diversity and Distributions}, 14(5):763--773.

\bibitem[Yu et~al., 2023]{yu2023future}
Yu, H., Wang, T., Skidmore, A., Heurich, M., and B{\"a}ssler, C. (2023).
\newblock How future climate and tree distribution changes shape the biodiversity of macrofungi across europe.
\newblock {\em Diversity and Distributions}, 29(5):666--682.

\end{thebibliography}

\end{document}